\documentclass{article}

\usepackage{arxiv}

\usepackage[utf8]{inputenc} 
\usepackage[T1]{fontenc}    
\usepackage{hyperref}       
\usepackage{url}            
\usepackage{booktabs}       
\usepackage{amsfonts}       
\usepackage{nicefrac}       
\usepackage{microtype}      
\usepackage{lipsum}		
\usepackage{graphicx}
\usepackage{doi}

\usepackage[labelformat=empty]{caption}
\usepackage{amsmath}
\usepackage{enumitem}
\PassOptionsToPackage{numbers, compress}{natbib}
\usepackage{natbib}

\newenvironment{packed_itemize}{
\begin{itemize}[noitemsep,topsep=0pt]
  \setlength{\itemsep}{1pt}
  \setlength{\parskip}{0pt}
  \setlength{\parsep}{0pt}
}{\end{itemize}}

\title{The Story of Airplane Wings}

\author{ Sam Greydanus \\
	The ML Collective\\
}
\date{} 

\begin{document}
\maketitle

\begin{abstract}
    \textit{When I was growing up, hummingbirds used to fly into our garage and get stuck. I remember finding one perched on a windowsill, weak from exertion. I carried it outside and opened my hands to release it, but it lay still for a moment. In that moment, the sunlight ignited its iridescent plumage and engulfed its whole body in a cloud of blues and greens. Then it understood it was free, whirred its wings, and vanished into the open air. Long after it had departed, my mind's eye gazed upon the little bird. With an idle curiosity, I tried to imagine how natural forces could have wrought such a thing. Since that encounter, I have always been naturally drawn to flight. It's a bizarre desire to possess, since humans did not evolve to fly. And yet many humans have felt the same way over the years. Entire cultures, even, have dreamed of flight. And step by step, they used technology to fashion their own artificial wings and bring about the modern-day miracle of flight.}\\

    In this work, we will recount that epic adventure. We will use the lens of history, looking at the individual people who wanted to fly, the lens of technology, looking at the key inventions leading up to modern airplanes, and the lens of physics, looking at the equations of airflow that made it all possible. Finally, we will derive our own wing from scratch. We will do this by simulating a wind tunnel, placing a rectangular occlusion in it, and then using gradient ascent to turn it into a wing.
\end{abstract}

\section{A History of the Early Aviators}

In the past century alone, airplane design has progressed from the Wright brothers’ ramshackle \textit{Flyer} to Lockheed Martin's streamlined \textit{SR-71 Blackbird}. In parallel, our commercial aviation system has developed to the point where anyone can enjoy the possibilities of flight. Flight has become so common and so reasonable that it's easy to forget the towering lusts and follies that accompanied its invention. But if we are to understand why humans wanted to fly in the first place, we need to relive those lusts and follies. One way to do this is by looking back at the lives of the early aviators.

Perhaps none of the early aviators wanted to fly as much as the tower jumpers. Beginning in the Middle Ages, there was a string of inventors who fashioned wings for themselves and then leaped from towers in imitation of birds. These daredevils were the tower jumpers. Having neglected the essential calculations of lift and drag, they relied mostly on wits, intuition, and dumb luck to survive. Alas, gravity tended to be stronger. Most of their attempts ended in death or serious injury. But one exception to the rule was the Andalusian inventor Abbas ibn Firnas. The story goes that he made his jump at the astonishing age of seventy years. Once he was airborne, his feather suit cushioned his fall and allowed him to glide to the ground unharmed \cite{white1961eilmer}. Somehow, in spite of his wild deeds, Abbas lived on to the respectable age of 78.

The common folk of Andalusia must have enjoyed laughter and endless conversations about Abbas and his wingsuit. They would have asked, why does he want to fly? Pigs do not have wings and they are quite content about it. Why should humans, who likewise have no wings, want to pretend at flight? They would have been making a good point. After all, the tower jumpers needed to be more than half mad to take the risks they did. But on a fundamental level, they were probably driven by things that most humans can relate to: the desires for freedom, glory, and adventure. 

We can see this in Leonardo da Vinci, whose work on flight was deeply rooted in a desire for freedom. Most people know that da Vinci painted the Mona Lisa and sketched a number of remarkable flying machines. However, few of them are aware of the pressures that shaped him \cite{isaacson2017leonardo}. Leonardo began life as the illegitimate son of an Italian aristocrat and a peasant woman. And so he had to support himself from an early age by taking on strict apprenticeships and painting for wealthy patrons. The fact that he was gay complicated things even further. His life reached a moment of crisis on his 27th birthday when he and his friends were imprisoned for acts of sodomy. Imprisonment affected him deeply. As soon as he was released, he sketched a machine meant to ``open a prison from the inside'' and another for tearing bars off of windows. Around the same time, he became obsessed with bird flight and began buying birds at markets in order to sketch them. When he finished sketching them, he would free them from their cages. ``Once you have tasted flight,'' he wrote, ``you will forever walk the earth with your eyes turned skyward. For there you have been, and there you will always long to return.'' For da Vinci, flight seems to have been a metaphor for freedom and an antidote to the horrors of captivity.

But freedom is not the only reason people wanted to fly. For others, flight was a shot at glory -- and nobody loved glory more than the French chemist Pilatre de Rozier. This man was quite a character. He was known to breathe fire, seduce older women, and give himself fake titles \cite{duval1967pilatre}. As a young scientist, he signed his papers, ``Apothecary,” then ``first Apothecary,” and finally, ``Pharmacy Inspector” of the Prince of Limbourg. This impressed his colleagues until they discovered that the Prince of Limbourg did not, in fact, exist. In a twist of irony, it was this same hunger for glory that also led de Rozier to his greatest scientific breakthroughs. The first was the match, which he invented in order to show off his fire breathing skills in a public lecture. The second was the gas mask, which he used in a stunt that involved lowering himself into the fumes of a vat of fermenting beer. The final, and most significant breakthrough, was his completion of the first manned voyage in a hot air balloon. King Louis XVI wanted to put criminals in the balloon but de Rozier objected, saying ``The glory should not be given to criminals!” \cite{penenberg2017sky}.

Pilatre was far from perfect, but his bravery and showmanship inspired people and gave aeronauts an immensely positive image in France. His initial balloon launch attracted a crowd of over a hundred thousand people. When these people saw him land safely, common attitudes about flight changed. No longer was human flight seen as a thing of folly. It became a source of national pride and a crowning achievement of science. And for many, it began to represent an exciting new frontier.

\begin{center}
\includegraphics[width=\textwidth]{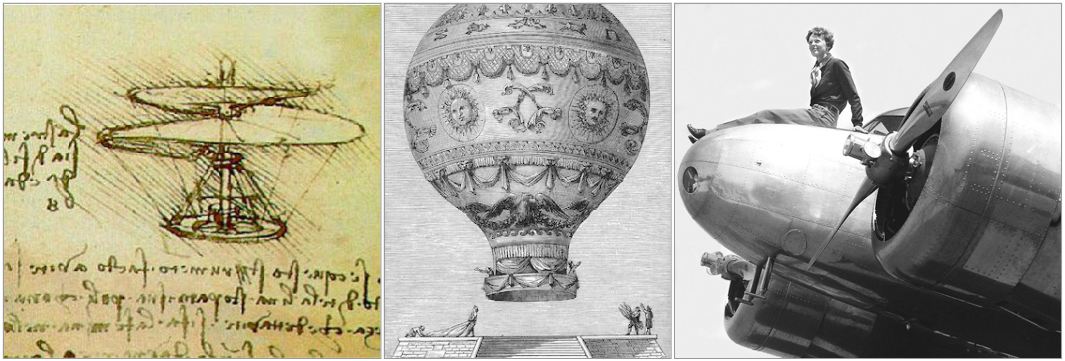}
\captionof{figure}{\textbf{Left:} A page from one of da Vinci's journals describes an air screw that resembles a modern helicopter. \textbf{Center:} Pilatre de Rozier and his companion take off in a hot air balloon. \textbf{Right:} Amelia Earhart sitting upon the nose of her plane in 1936.}
\end{center}

Like any new frontier, flight called most strongly to those with a sense of imagination and adventure. One person who heeded this call was Antoine de Saint-Exupery. In his short life, he became both a decorated military pilot and one of France's best poets and novelists. He managed this double act by blending his writing with his flying until they became almost the same activity. There are eccentric stories about how he wrote poetry during military scouting missions and once circled a landing strip for an hour to finish reading a good novel \cite{schiff1997saint}. To Saint-Exupery, flight was as much an adventure of the mind as it was the body. In fact, he argued that these two adventures were inseparable: the way a mind perceives the physical world determines how the body will eventually shape it. Or, as he put it, ``A rock pile ceases to be a rock pile the moment a single man contemplates it, bearing within him the image of a cathedral.'' His own life supports that idea, for, in thinking and writing about flight from the lens of an artist, he transformed it into something that promised not only freedom and glory, but also beauty and adventure.

The other side of venturing into a frontier is that one must leave behind the comforts of civilized life. Even as airplanes improved in the early 1900's, they remained costly, dangerous, and generally impractical. So every one of the early aviators had to make sacrifices in order to fly. But one person who had to make especially tough choices was Amelia Earhart. Earhart was one of the best early airplane pilots and famously went missing during an attempt to fly around the world. The first, and most obvious choice she had to make was to risk her life in pursuit of that goal. But apart from safety, she had to risk bankruptcy throughout her twenties -- taking odd jobs and once selling her plane in order to support herself. And finally, once she was able to support herself financially, she had to risk losing love and the chance at marriage. The publicist George Putnam had proposed and the two were preparing for marriage when she explained to him, ``I cannot endure at all times even an attractive cage'' \cite{earhart2005prenup}. Putnam eventually agreed to an unconventional open marriage, but Earhart's letter suggests she was prepared to forgo it entirely if it interfered with her ability to fly. Like Earhart, many of the early aviators had to lose wealth, love, or life in the pursuit of flight. That was the nature of life on the new frontier.

\section{The Technological Stepping Stones of Flight}

By the year 1900, humans seemed to have figured out all the important ideas of aviation. There were patents for dozens of self-powered aircrafts including biplanes \cite{wenham1866aerial}, seaplanes \cite{allward1997illustrated}, and bombers \cite{crosby1997guide}. There were also designs for retractable landing gear \cite{gibbs2003aviation}, aileron wing controls, and curved airfoils \cite{boulton1864locomotion}. To many people, these designs indicated that the age of the airplane was at hand. Indeed, by this time there were already two commercial airline startups \cite{national1843ariel} and one program for constructing bombers, fully funded by the French government \cite{crosby1997guide}. But there was just one problem: nobody had managed to fly a real airplane yet. In fact, the Wright brothers were still several years from achieving that breakthrough. As for bombers, seaplanes, and airline companies -- such things were still many decades away.

This bizarre gap between theory and practice brings into question the meaning of the word invention. Generally speaking, one would think of an invention as a detailed design of the sort that could be patented. But in the case of the airplane, dozens of people patented airplanes that never could have flown. Did those people really invent airplanes? Otto Lilienthal, the first glider pilot, would have answered in the negative. ``To design an aircraft is nothing” he wrote, ``To build one is something. But to fly is everything.”

\begin{center}
\includegraphics[width=0.3\textwidth]{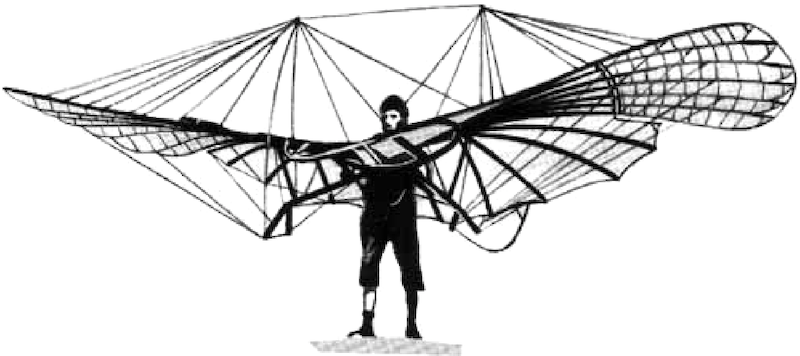}
\captionof{figure}{Otto Lilienthal}
\end{center}

In order to focus on the practical stepping stones of flight, let's adopt Lilienthal's perspective. In other words, let's focus on the changes in wing design that led to the most significant empirical improvements in flight. When we start down this path, one of the first things we notice is that most of these improvements are related to the cross-sectional shape of a wing, also known as the airfoil. Generally speaking, an airfoil has no moving parts. In fact, it is just a two-dimensional shape that influences the speed of air above and below a wing. Prior to the Wright brothers, few people gave serious thought to airfoil design. But it just so happened that the details of this shape play a critical role in determining the lift and drag profile of a wing. Only through repeated iteration of design and demonstration did we become aware of the airfoil's surprising complexity.

Lilienthal, with his emphasis on practical results, was the first person to realize how much airfoils matter. They came to his attention when he set out to build a glider that could support the weight of a man. In order to build such an apparatus, he needed to know how the lift and drag characteristics of various wings scaled with length, width, and thickness. In thinking about these questions, he quickly realized that the cross-sectional shape was an important design parameter. He started by studying the airfoils of bird wings, especially those of storks. Then, with their shapes in mind, he built artificial replicas in his laboratory and tested their lift coefficients. These methodical experiments took nearly twenty years of work, after which he published his findings in his thorough (and beautifully illustrated) book, \textit{Birdflight as the Basis of Aviation} \cite{lilienthal1889birdflight}. Taking what he had learned from biology, Lilienthal spent the latter part of his life building and flying human-scale gliders -- the first winged flying machines that could carry a human through the air. And so, with an eye on nature and great attention to detail, Lilienthal achieved the goal he had been working towards since his twenties.

But even Lilienthal, for all his care, made some errors. The Wright brothers used his results to design their first glider, and when that glider crashed unexpectedly, they realized that something was wrong. After performing their own set of wind tunnel experiments, they determined that the widely-accepted value of the Smeaton coefficient -- a key part of Lilienthal's lift and drag equations -- was about 60\% too high.\footnote{The wrong value of this coefficient had been in use for more than a hundred years and was part of the accepted equation for lift. So in determining that this number was wrong, the Wrights made a major discovery.} Starting from this discovery, they reworked their entire wing design. They tested out hundreds of airfoils in a miniature wind tunnel and finally settled on a new airfoil shape. It was slightly wider and more arched, and the highest point of its arch was closer to the front of the wing (more ``forward camber''). In spite of these changes, it looked only slightly different from their original wing shape. And yet its improved lift and stability was what the Wrights needed in order to build the world's first self-propelled airplane. 

\begin{center}
\includegraphics[width=\textwidth]{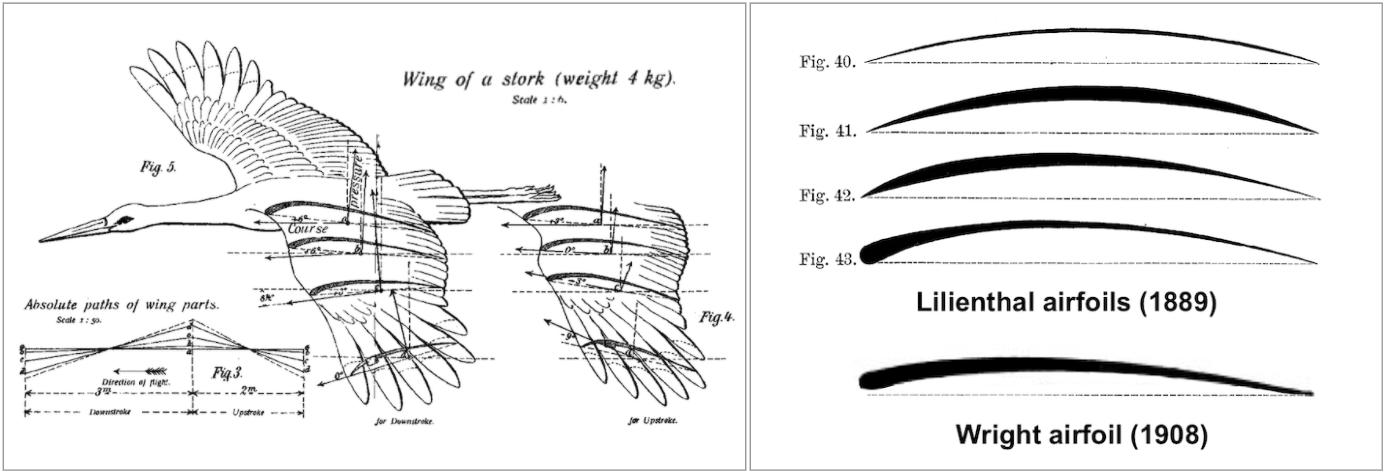}
\captionof{figure}{\textbf{Left:} A figure from Otto Lilienthal's book \textit{Birdflight as the Basis of Aviation}. Lilienthal studied bird wings and carried out small scale experiments for two decades (1867-1889) in preparation for building his gliders. \textbf{Right:} Years of careful experiments by the Wright brothers yielded a new wing shape with a slightly more forward camber. This seemingly small change was crucial to the \textit{Flyer's} success.}
\end{center}

Although Lilienthal and the Wrights lived on different continents and had very different lives, they were united by the fact that they each spent years doing tedious, small-scale experiments in order to understand flight on a deep level. Only after this exploratory phase did they build real flying machines. In a way, those long hours of tedium are a sacrifice one must make when they set out to pursue the romantic ideal of flight. In profiling the early aviators, we saw that the frontier of flight often attracted radical and temperamental dreamers. These were not reliable people. And yet each of them had to discipline and civilize themselves, becoming the most practical of the lot of us, before they could become heroes. Interestingly, the next breakthroughs in flight were to involve the same dynamic, but this time playing out across nations rather than individuals.

These breakthroughs were the product of national labs. National labs were a new phenomenon that emerged in the 1910's and 1920's as forward-thinking governments started to reckon with the military and economic applications of flight. Leading up to World War I, airplanes were still slow, unreliable, and expensive. They were great for stunts and parades, but close to useless on the battlefield. One of the main goals of the national labs was to change this.

The first improvements in airfoil design came out of Britain's National Physical Laboratory. Managers at this lab used its infrastructure and manpower to test airfoil designs much more thoroughly than the hobbyist-inventors before them. One of their key findings was that thicker wings with even more forward camber gave better lift. As soon as they made this discovery, they built it into military planes like the Airco DH.2 fighter (1915) and the Vickers Vimy bomber (1917).

But even though these changes in airfoil design improved lift, they did not improve stability. World War I era biplanes suffered from a dangerous effect called ``thin airfoil stall.'' This effect occurred when streams of air above and below a wing collided behind it, creating unpredictable drag and sending the plane into a stall. German engineers were the first to find a solution to this problem. They found that thicker airfoils like the Göttingen 398 could mitigate thin airfoil stall and make fighter planes more maneuverable. They used these insights to build the Fokker Dr. I (1917) which was one of the most dangerous fighters of the war \cite{topnotch2020history}.

The most amazing thing about flight research during World War I was the speed at which national labs turned research into real technology. Often it only took a year or two. The condensed timeline and extreme real-world impact of airplanes led to dramatic improvements in their designs and finalized their transition from the world of ideas to the world of things. And the wonderful thing about having an idea take root in the world is its tendency to become the bedrock for an entirely new generation of ideas.

That is the story of the 1920's and 1930's, which is when the mathematical theory of flight got started. Government physicists in the United States finally had time to come up with theories that explained experimental results. Then they used these theories to make airfoils better in small but important ways. Their work culminated in the 1933 National Advisory Committee for Aeronautics (NACA) Report 460, which set the industry standard for the next several decades. World War II planes like the DC-2 transport and the B-17 Flying Fortress used these results \cite{century2020airfoils}. And after the war, designs like the NACA 2412 found their way into commercial plane designs, some of which are still in use today.

\begin{center}
\includegraphics[width=\textwidth]{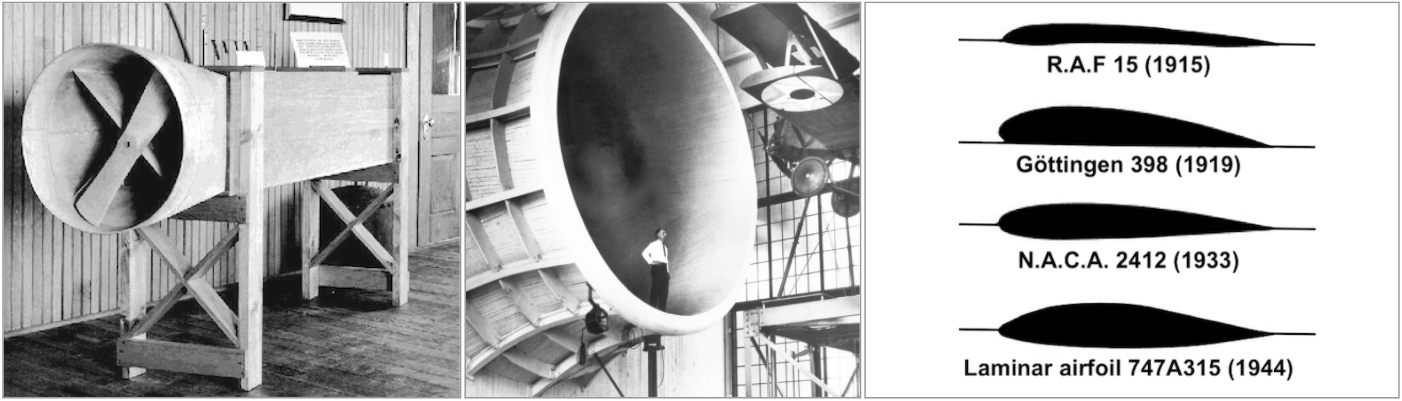}
\captionof{figure}{\textbf{Left:} The 1901 wind tunnel that the Wright brothers used to study airfoil shapes. \textbf{Center:} A man standing in the enormous Langley wind tunnel in 1925. \textbf{Right:} The progression of notable airfoils developed by national labs between 1915 and 1945.}
\end{center}

The process of minor improvements based on theory continued into the 1940's, when NACA researchers invented the laminar flow airfoil and installed it on the P-51 Mustang \cite{century2020airfoils}. In practice, the laminar flow ``correction'' was rather small and it led to modest improvements. But it represented a milestone in that it was one of the first major innovations motivated by theory rather than human intuition or observations of biology. Focusing on the causal mechanisms of flight ended up being crucial for later innovations in the supersonic regime since the way air behaves at those speeds is much less intuitive.

In fact, there is a deep connection between how well we understand the laws of nature and what we can build in the world. The laws of nature are the rules of the game. We are constantly learning more about these rules and we can only innovate in proportion to how well we understand them. To see this, consider evolution for a moment. Over millions of years, it has deformed life so as to probe the laws of nature at many different scales. So with time, the fundamental forces of nature have constrained and shaped life into the variety of forms we see today. Human design mimics this trial-and-error approach, but our mental models of the world give us an advantage. They speed our search in proportion to how much of the physical world they can explain. And by acting on our mental models, we can make intuitive leaps that evolution, in all of its billions of years, never could have managed. One such intuitive leap was made by Richard Whitcomb when he discovered supercritical airfoils.

This discovery occurred in 1965, which was a time when the aerospace industry was trying to improve supersonic flight. Jet engines, which were invented at the end of World War II, had improved to the point where they produced enough force to accelerate planes to supersonic speeds. But once planes reached these speeds, the physics of airflow started to change and existing airfoil designs stopped working. NACA researchers realized that they would have to rethink every aspect of wing design in order to adapt. One of the most challenging problems was what to do with airfoil design \cite{century2020airfoils}.

At the time, many of Whitcomb's colleagues were looking for solutions in aerodynamic theory. Whitcomb took a different approach: he grabbed a can of putty and headed for the Langley wind tunnel \cite{garrison2005man}. He knew that the problem with existing airfoils was that air flowed at a higher rate around the top of the wing than the bottom. As the plane approached supersonic speeds, the air on top was the first to hit the sound barrier. Energy, normally dissipated as sound, would then be moving at the same speed as the air itself and slowly start to accumulate. A shock wave would form. Then the shock wave would create all sorts of pathological drag and instabilities \cite{nasa1986airfoils}.

\begin{center}
\includegraphics[width=\textwidth]{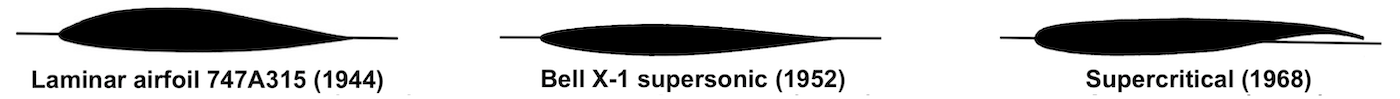}
\end{center}

With this in mind, Whitcomb used putty to decrease the camber of the wing so as to lower the airspeed above it. Then he added a slight concavity to the underside of the wing to maintain lift and stability. All of this was based on his intuition for how air flowed over a wing at high speeds, but it ended up being extraordinarily effective. His ``supercritical'' wing design allowed the United States to build the fastest bombers, fighters, and reconnaissance planes of the Cold War. And surprisingly, this wing design turned out to be stable and efficient even at subsonic speeds. Today's commercial airliners, which cruise at speeds around Mach 0.85, all use supercritical airfoils to improve fuel efficiency.

Looking over the history of wing design, it is easy to see that the boundary between imagination and the constraints of the real world is where invention happens. When ideas are fully constrained to our minds, we have the tendency to indulge in impractical fantasies. And yet we need imagination too. For without it, we are limited to the incremental trial-and-error pace of evolution. Imagination is our one clear advantage over evolution, for it requires no intermediary. For evolution to invent a wing, there needed to be a half-winged precursor. But imagination has a strangely liberating effect in that it allows us to move from the ground to the sky in a single intuitive leap.

\begin{center}
\includegraphics[width=\textwidth]{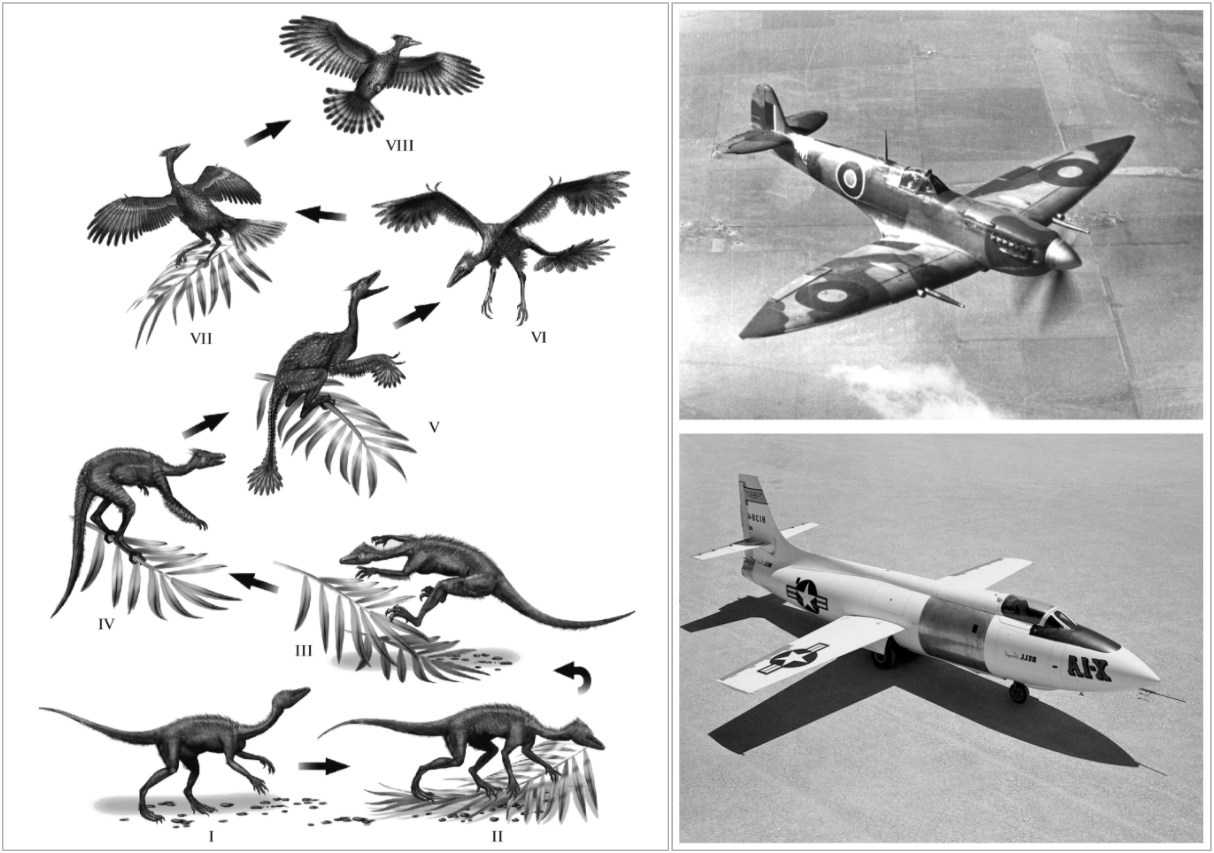}
\captionof{figure}{\textbf{Left:} An evolutionary pathway from small sauropods to flying birds, proposed by Kurochkin and Bogdanovich (2010) \cite{kurochkin2016origin}. Each intermediary took hundreds of thousands of years of natural selection. \textbf{Right:} The Supermarine Spitfire (top) was one of the fastest planes of World War II. The Bell X-1 (bottom) unseated the Spitfire and broke the sound barrier a few years later. Notice how different the two planes look; human design lacks intermediaries.}
\end{center}

Since the first half of the twentieth century when the core breakthroughs of aeronautics occurred, scientists have been hard at work on physical theories that can explain those breakthroughs. These theories have expanded into fields of study like ``computational fluid dynamics,'' ``aeronautical science'', and ``turbulent flow.'' Such principles are quite complex, but the question they aim to answer is simple: \textit{how do wings work?} We are going to answer that question in the next section by obtaining our own wing starting from nothing but the physics of airflow.

\section{Using the Physics of Airflow to Optimize a Wing}

Legos are an excellent meta-toy in that they represent the potential for a near-infinite number of toys depending on how you assemble them. Each brick has structure. But each brick is only interesting to the extent that it can combine with other bricks, forming new and more complex structures. So in order to enjoy Legos, you have to figure out how they fit together and come up with a clever way of making the particular toy you have in mind. Once you have mastered a few simple rules, the open-ended design of Lego bricks lets you build anything you can imagine.

Our universe has the same versatile structure. It seems to run according to just a few simple forces, but as those forces interact, they give rise to intricate patterns across many scales of space and time. You see this everywhere you look in nature -- in the fractal design of a seashell or the intricate polities of a coral. In the convection of a teacup or the circulation of the atmosphere. And this simple structure even determines the shape and behavior of man's most complicated flying machines.

To see this more clearly, we are going to start from the basic physical laws of airflow and use them to derive the shape of a wing.\footnote{Specifically, we build on ideas laid out in \citet{maclaurin2015autograd}.} Since we are using so few assumptions, the wing shape we come up with will be as fundamental as the physics of the air that swirls around it. This is pretty fundamental. In fact, if an alien species started building flying machines on another planet, they would probably converge on a similar shape.

We will begin this journey with the Navier-Stokes equation, which sums up pretty much everything we know about fluid dynamics. It describes how tiny fluid parcels interact with their neighbors. The process of solving fluid dynamics problems comes down to writing out this equation and then deciding which terms we can safely ignore. In our case, we would like to simulate the flow of air through a wind tunnel and then use it to evaluate various wing shapes.

Since the pressure differences across a wind tunnel are small, one of the first assumptions we can make is that air is incompressible. This lets us use the incompressible form of the Navier-Stokes equation:
\begin{equation*}
    \underbrace{\frac{\partial \mathbf{u}}{\partial t}}_{\text{velocity update}} ~=~ - \underbrace{(\mathbf{u} \cdot \nabla)\mathbf{u}}_{\text{self-advection}} ~+~ \underbrace{\nu \nabla^2 \mathbf{u}}_{\text{viscous diffusion}} ~+~ \underbrace{f}_{\text{velocity $\uparrow$ due to forces}}
\end{equation*}
Another term we can ignore is viscous diffusion. Viscous diffusion describes how fluid parcels distribute their momenta due to sticky interactions with their neighbors. We would say that a fluid with high viscosity is ``thick": common examples include molasses and motor oil. Even though air is much thinner, viscous interactions still cause a layer of slow-moving air to form along the surface of an airplane wing. However, we can ignore this boundary layer because its contribution to the aerodynamics of the wing is small compared to that of self-advection.

The final term we can ignore is the forces term, as there will be no forces on the air once it enters the wind tunnel. And so we are left with but a hair of the original Navier-Stokes hairball:
\begin{equation*}
    \underbrace{\frac{\partial \mathbf{u}}{\partial t}}_{\text{velocity update}} = \underbrace{- (\mathbf{u} \cdot \nabla)\mathbf{u}}_{\text{self-advection (``velocity follows itself")}}
\end{equation*}
This simple expression describes the effects that really dominate wind tunnel physics. It says, intuitively, that ``the change in velocity over time is due to the fact that velocity follows itself.'' So the entire simulation comes down to two simple rules:

\underline{Rule 1: Velocity follows itself.} The technical term for this effect is ``self-advection.'' Advection is when a field, say, of smoke, is moved around by the velocity of a fluid. Self-advection is a special case where the field being advected is the velocity field, and so it actually advects itself. In principle, a self-advection step is as simple as moving the velocity field according to $x_1 = v_0 \Delta t + x_0$ at every point on the grid. We can simulate self-advection over time by repeating this over and over again.\footnote{We'd call this an ``Euler integration'' of the dynamics. The problem with Euler integration is that when you run it on a grid, small numerical errors can accumulate into big ones. There's a related approach called the ``Backward Euler'' method which mitigates these errors. In Backward Euler, we use the final velocity rather than the initial velocity to perform advection: the update becomes $x_1 = v_1 \Delta t + x_0$ instead. To gain deeper intuition for why this is a good idea, refer to page eight of \citet{stam2003real}.}
    
\underline{Rule 2: Volume is conserved.} This rule comes from our incompressibility assumption and the process of enforcing it is called projection. Since volume is conserved, fluid parcels can only move into positions that their neighbors have recently vacated. This puts a strong constraint on our simulation's velocity field: it needs to be volume-conserving. Fortunately, Helmholtz’s theorem tells us that any vector field can be decomposed into an incompressible field and a gradient field, as this figure from \cite{stam2003real} shows:
\begin{center}
    \includegraphics[width=.5\textwidth]{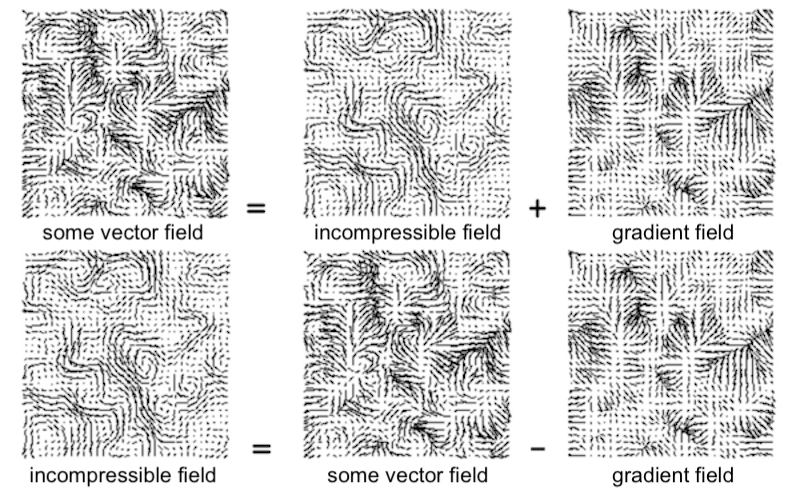}
\end{center}
One way to make our velocity field incompressible is to find the gradient field and then subtract it from the original field as shown above\footnote{Following \citet{stam2003real}, we implement this step by using a few iterations of the Gauss-Seidel method to solve Poisson's equation.}. This \textit{projects} our velocity field onto a volume-conserving manifold.

By alternating between these two rules, we can iteratively 1) move the system forward in time and 2) enforce conservation of volume and mass. In practice, we implement each rule as a separate function and then apply both functions to the system at every time step. This allows us to simulate, say, a gust of wind passing through the wind tunnel. But before we can direct this wind over a wing, we need to decide how to represent the wing itself.

The wing is an internal boundary, or occlusion, of the flow. A good way to represent an occlusion is with a mask of zeros and ones. But since the goal of our wind tunnel is to try out different wing shapes, we need our wing to be continuously deformable. So we will allow the mask to take on continuous values between zero and one, making it semi-permeable in proportion to its mask values\footnote{In practice, the wing is still not quite continuously deformable. Big differences in the mask at neighboring grid points can lead to sharp boundary conditions and non-physical airflows around the mask. One way to reduce this effect is to apply a Gaussian filter around the edges of the mask so as to prevent these grid-level pathologies. Although this may seem like an arbitrary design decision at first glance, it is actually a common technique used in a wide range of other physics simulations such as topology optimization \cite{sigmund200199}, large eddy simulation \cite{bose2010grid, gullbrand2003grid}, and 3D graphics \cite{levin2004mesh}.}. This lets us add semi-permeable obstructions to the wind tunnel as shown:
\begin{center}
\includegraphics[width=\textwidth]{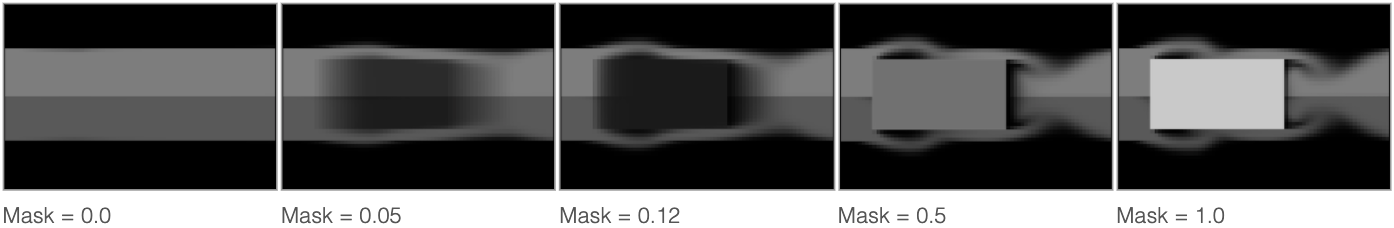}
\end{center}
Now we are at the point where we can simulate how air flows over arbitrary, semi-permeable shapes. But in order to determine which of these shapes makes a better wing, we still need to define a measure of performance. There are many qualities that one could look for in a good wing, but we will begin with the most obvious: it should convert horizontal air velocity into upward force as efficiently as possible. We can measure this ability using something called the lift-drag ratio where "lift" measures the upward force generated by the wing and "drag" measures the frictional forces between the air and the wing. Since "change in downward airflow" in the tunnel is proportional to the upward force on the wing, we can use it as a proxy for lift. Likewise, "change in rightward airflow" is a good proxy for the drag forces on the wing. With this in mind, we can write out the objective function as
\begin{equation*}
    \max_{\theta} L/D
\end{equation*}
where $\theta$ represents some tunable parameters associated with the shape of the wing mask and $L/D$ can be obtained using the initial and final wind velocities of the simulation according to
\begin{align}
     L/D &= \frac{\text{lift}}{\text{drag}}\\
    &= \frac{\text{change in downward airflow}}{-\text{change in rightward airflow}}\\
    &= \frac{ -\big ( v_y(t)-v_y(0) \big )}{-\big ( v_x(t)-v_x(0) \big )}\\
    &= \frac{ v_y(t)-v_y(0) }{ v_x(t)-v_x(0)}
\end{align}
Solving this optimization problem will give us a wing shape that generates the most efficient lift possible. In other words, we new have the correct problem setup; what remains is to figure out how to solve it.

We are going to solve this problem with gradient ascent. Gradient ascent is simple and easy to implement, but there is one important caveat: we need a way to efficiently compute the gradient of the objective function with respect to the wing mask parameters. This involves differentiating through each step of the fluid simulation in turn -- all of the way back to the initial conditions. This would be difficult to implement by hand, but fortunately there is a tool called Autograd which can perform this back-propagation of gradients automatically\footnote{Amazingly, every mathematical operation we've described so far – from the wing masking operation to the advection and projection functions to the lift-drag ratio – is differentiable. This is why we can use Autograd to compute analytic gradients with respect to the mask parameters. Autograd uses automatic differentiation, closely related to the adjoint method, to propagate gradient information backwards through the simulation until it reaches the parameters of the wing mask. We can do all of this in a one-line function transformation: \texttt{grad\_fn = autograd.value\_and\_grad(get\_lift\_drag\_ratio)}.}. We will use Autograd to compute the gradients of the mask parameters, move the mask parameters in that direction, and then repeat this process until the lift-drag ratio reaches a local maximum.

So let’s review. Our goal is to simulate a wind tunnel and use it to derive a wing shape. We began by writing down the general Navier-Stokes equation and eliminating irrelevant terms: all of them but self-advection. Next, we figured out how to represent a wing shape in the tunnel using a continuously-deformable occlusion. Finally, we wrote down an equation for what a good wing should do and discussed how to optimize it. Now it is time to put everything together in about two hundred lines of code and see what happens when we run it...
\begin{center}
\includegraphics[width=\textwidth]{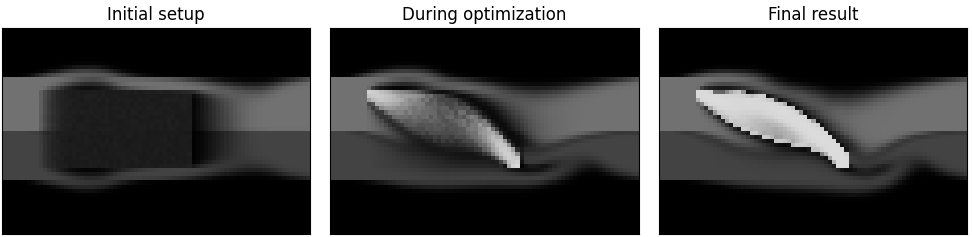}
\end{center}
Sure enough, we get a beautiful little wing. Of all possible shapes, this is the very best one for creating efficient lift in our wind tunnel. This wing is definitely a toy solution since our simulation is coarse and not especially accurate. However, after making a few simple improvements we would be able to design real airplane wings this way. We would just need to:
\begin{packed_itemize}
    \item Simulate in 3D instead of 2D
    \item Use a mesh parameterization instead of a grid
    \item Make the flow laminar and compressible
\end{packed_itemize}
Aside from these improvements, the overall principle is much the same. In both cases, we write down some words and symbols, turn them into code, and then use the code to shape our wing \cite{northwestern2020optimization}. The fact that we can do all of this without ever building a physical wing makes it feel a bit like magic. But this process really works, for when we put these wings on airplanes and trust them with our lives, they carry us safely to our destinations \cite{jameson2001computational, jameson2010airplane}.

Just like the real wind tunnels of the twentieth century, these simulated wind tunnels need to go through lots of debugging before we can trust them. In fact, while building this demo we discovered a number of ways that things can go wrong. Here are some of the most amusing failure cases:
\begin{center}
\includegraphics[width=\textwidth]{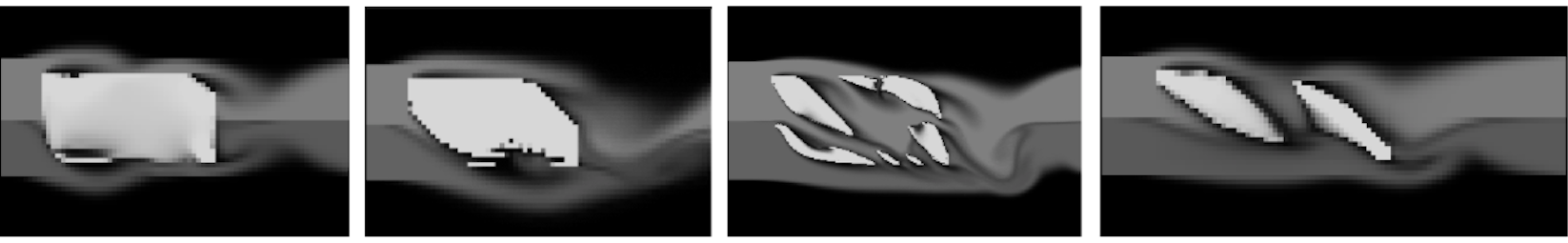}
\end{center}
Several of these wings are just plain dreadful. But others seem reasonable, if unexpected. The two-wing solution is particularly amusing. We did not intend for this ``biplane'' solution to occur, and yet it is a completely viable way of solving the objective we wrote down. One advantage to keeping the problem setup so simple is that, in doing so, we left space for these surprising behaviors to occur.

In fact, even though we stressed that the shape of our wing is as fundamental as the physics of the air that swirls around it, there are a number of variations on that shape which excel in particular niches. Sometimes we will want a wing that is optimal at high speeds and other times we will want one that is optimal at low speeds. In order to accommodate a large fuselage, we might want an extra-thick wing. Alternatively, in order to reduce its overall weight, we might want to keep it thin. It turns out that we can change simulation parameters and add auxiliary losses to find optimal wing shapes for each of these scenarios.

\begin{center}
\includegraphics[width=0.7\textwidth]{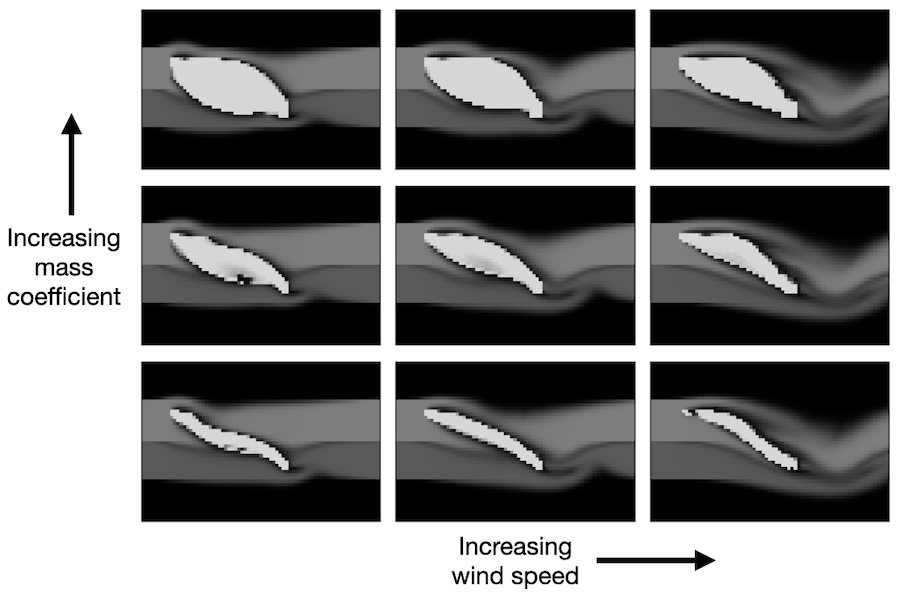}
\end{center}

Our wind tunnel simulation is interesting first, because it illustrates how the Platonic ideal of wing design is rooted in the laws of physics. As we saw in the first part of this essay, there were many cultural and technological forces that contributed to airfoil design. These forces were important for many reasons, but they were not the primary factor in the wing shapes they produced -- physics was.

But to balance this idea, we have also shown how a million variants of the Platonic form of a wing can fulfill particular needs. Indeed, these variants could be said to occupy complimentary niches in the same way that different birds and flying insects occupy different niches in nature. After all, even though nature follows the laws of physics with absolute precision, it takes a consummate joy in variation. Look at the variety of wing shapes in birds, for example. Species of hummingbirds have wings with low aspect ratios that enable quick, agile flight patterns. Other birds, like the albatross, have high aspect ratios for extreme efficiency. Still others, like the common raven, are good all-around fliers. Remarkably, we are beginning to see this same speciation occur in modern aircraft as well. There are surveillance planes built for speed and stealth, short-winged bush planes built for maneuverability, and massive commercial airliners built for efficiency.

\begin{center}
\includegraphics[width=\textwidth]{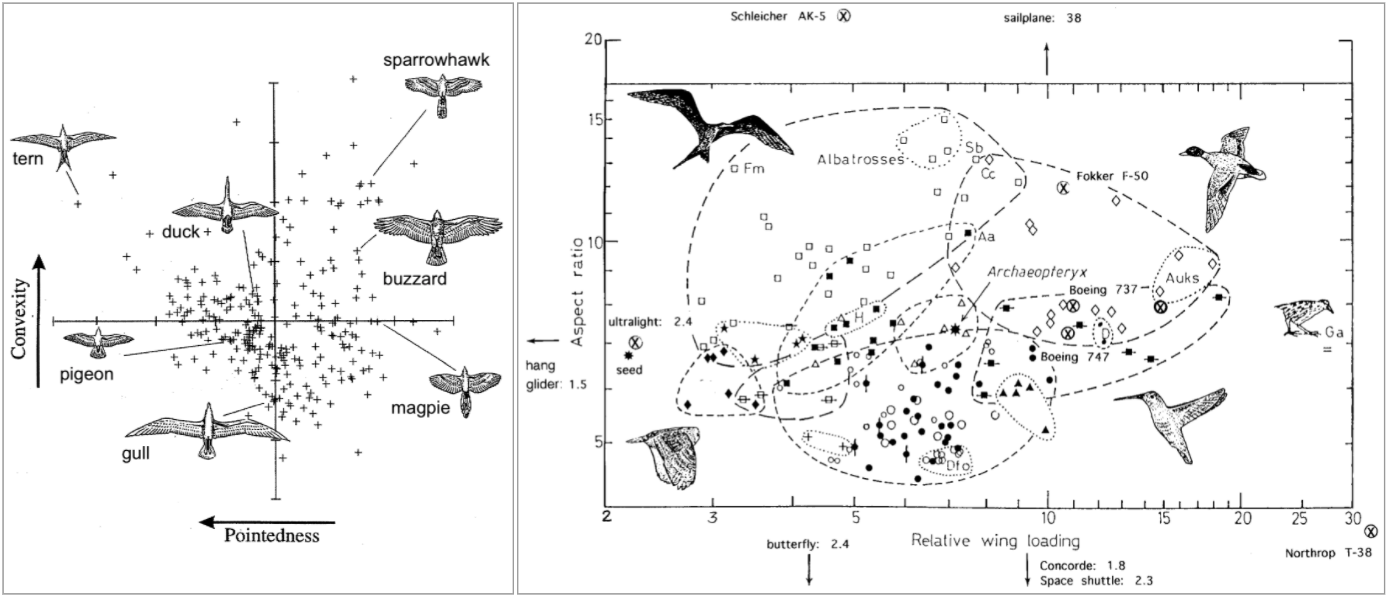}
\captionof{figure}{\textbf{Left:} A figure from \citet{lockwood1998avian} arranging bird species by wing pointedness and wingtip convexity. Different wing designs stem from adaptations to different ecological niches. \textbf{Right:} A plot showing aspect ratio versus wing loading index in some birds, airplanes, a hang-glider, a butterfly, and a maple seed \cite{lindhe2002structure}. Just like the families of birds, different human flying machines display substantial variation along these axes.}
\end{center}

Perhaps less intuitively, even a single bird is capable of a huge range of wing shapes. The falcon, for example, uses different wing shapes for soaring, diving, turning, and landing. Its wings are not static things, but rather deformable, dynamic objects which are constantly adapting to their surroundings. And once again, we are beginning to see the same thing happen in modern aircrafts like the Boeing 747. The figure below shows how its triple-slotted wing design lets pilots reconfigure the airfoil shape during takeoff, cruising, and landing.

\begin{center}
\includegraphics[width=\textwidth]{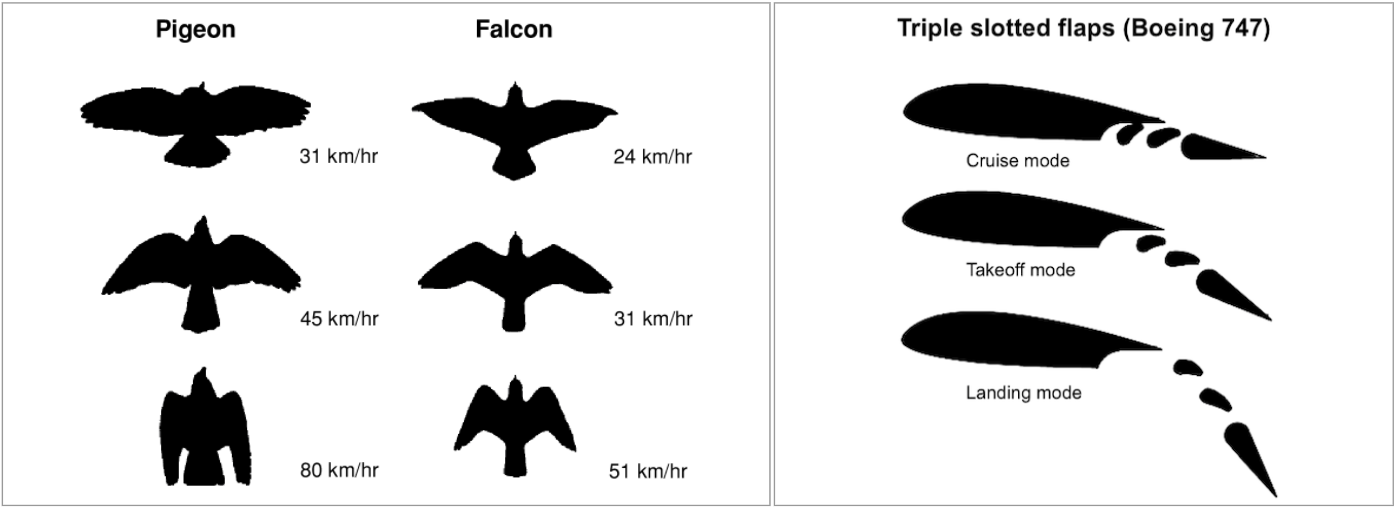}
\end{center}

One of the lessons from attempting to optimize a wing is that the optimization itself is never the full story. When we write down the optimization objective (like we did above), our minds already have a vague desire to obtain a wing. And behind that desire, our minds may want to obtain a wing because we are drawn to the technology of flight. And perhaps we are drawn to flight for the same reasons that the early aviators were -- because it promises freedom, glory, and adventure. And behind those desires -- what? The paradox of an objective function is that it always seems to have a deeper objective behind it.

The deeper objectives do not change as quickly. Even as the early aviators progressed from wingsuits to gliders to planes, they retained the same fundamental desire to fly. Their specific desires, of course, were different: some wanted to survive a tower jump and others wanted to break the speed of sound. And their specific desires led to specific improvements in technology such as a better understanding of the Smeaton coefficient or a more stall-resistant airfoil. Once they made these improvements, the next generation was able to use them to pursue more ambitious goals. But even as this cycle progressed, the desire to fly continued to inspire and even unify their efforts.



\section*{Acknowledgements}

Thanks to \citet{maclaurin2015autograd} for releasing Autograd to the world along with a number of thought-provoking demos. Thanks to Stephan Hoyer, Shan Carter, and Matthew Johnson for conversations that shaped some of the early versions of this work. And thanks to Andrew Sosanya, Jason Yosinski, and Tina White for feedback on early versions of this essay. Special thanks to my family and friends for serving as guinea pigs for early iterations of this story.

\bibliographystyle{plainnat}
\bibliography{references}

\newpage 
\appendix

\begin{center}
    \textsc{\LARGE \bf Appendix}
\end{center}

Here is a complete timeline of the airfoils we discussed in this work. Outlines in the first three columns were obtained from primary sources and are technically accurate. Outlines in the fourth row are believed to be technically accurate.
\begin{center}
\includegraphics[width=\textwidth]{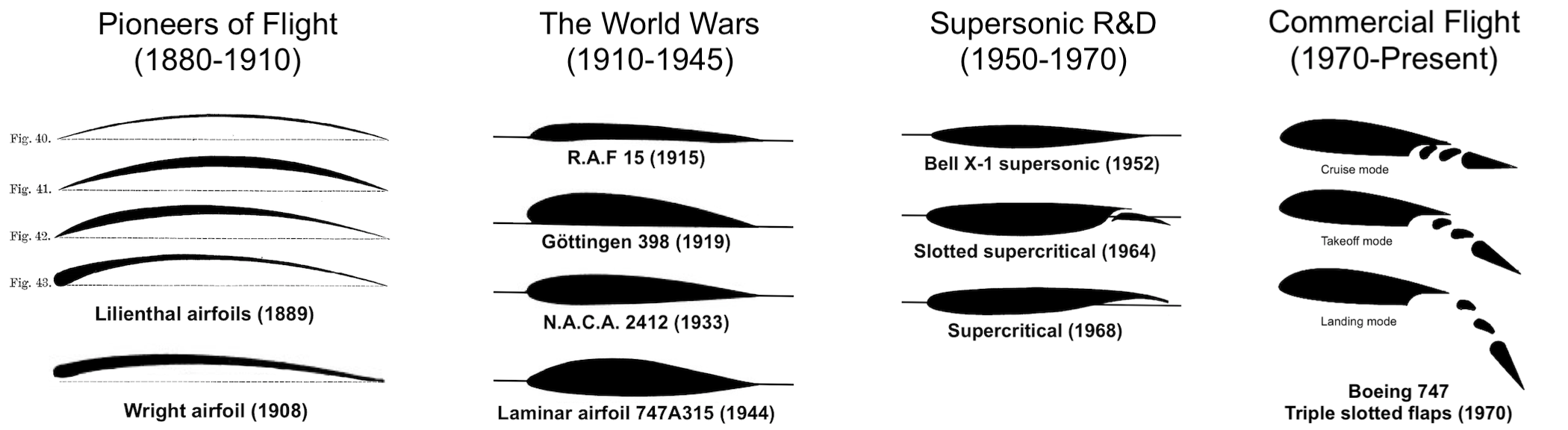}
\end{center}

\end{document}